\documentclass[cameraready]{Interspeech}

\title{ZEBRA: Zero-Shot Entropy-Regularized Prompt Learning for Base-to-Novel Generalization in Audio-Language Models}

\author[orcid=0000-0002-6705-149X,correspondingauthor]{Asif}{Hanif}
\author[orcid=0000-0001-6896-1105]{Mohammad}{Yaqub}

\address{
    Mohamed Bin Zayed University of Artificial Intelligence, Abu Dhabi, UAE
}

\email{asif.hanif@mbzuai.ac.ae, mohammad.yaqub@mbzuai.ac.ae}


\usepackage[dvipsnames]{xcolor}

\newcommand{\x}{\bm{\mathrm{x}}}
\newcommand{\txt}{\bm{\mathrm{t}}}

\usepackage{pifont}
\newcommand{\cmark}{\ding{51}}%
\newcommand{\xmark}{\ding{55}}%

\newcommand{\uptri}[1]{\footnotesize\textcolor{gray}{(\textcolor{ForestGreen}{$\blacktriangle$}{#1})}}
\newcommand{\downtri}[1]{\footnotesize\textcolor{gray}{(\textcolor{BrickRed}{$\blacktriangledown$}{#1})}}

\newcounter{daggerfootnote}

\usepackage{marvosym}

\begin{document}

\maketitle

\begin{abstract}
Audio-Language Models (ALMs) achieve strong zero-shot performance by aligning audio with textual class descriptions. Although prompt learning improves accuracy on base classes through few-shot supervised adaptation, we observe a critical trade-off: it often degrades performance on novel classes, sometimes falling below zero-shot accuracy. This exposes a base-to-novel generalization gap in prompt learning for ALMs. To address this issue, we propose \textbf{ZEBRA} (Zero-shot Entropy-Regularized Prompt Learning for Base-to-Novel Generalization), a plug-and-play framework that fuses zero-shot logits with prompt-learning logits, and employs self-entropy regularization to reduce overfitting to base classes. Experiments across multiple audio classification datasets show that  ZEBRA consistently improves novel-class performance while maintaining strong base accuracy, significantly reducing the base-to-novel gap compared to standard prompt learning. The code is available at: \url{https://github.com/asif-hanif/zebra}.

\end{abstract}

\section{Introduction}
\label{sec:intro}
Recent advances in Vision-Language Models (VLMs) have inspired the development of Audio-Language Models (ALMs), which achieve strong performance on zero-shot audio recognition tasks \cite{deshmukh2023pengi,elizalde2023clap,guzhov2022audioclip,zhang2024vision}. In the zero-shot setting, audio features are aligned with textual descriptions of class labels, enabling recognition without task-specific training. This paradigm provides flexibility and generalization to unseen categories while eliminating the need for large labeled datasets. Despite these advantages, zero-shot performance often falls short of fully supervised or few-shot adaptation. The primary limitation lies in the reliance on manually designed text prompts, which may not optimally align with the target dataset. Prompt learning addresses this limitation by replacing hand-crafted templates with learnable context tokens that are optimized using labeled data from base classes \cite{zhou2022coop,zhou2022conditional}. By adapting prompts to the downstream task, prompt learning significantly improves performance over zero-shot baselines.\\

\begin{figure}
    \centering
    \includegraphics[width=1\linewidth]{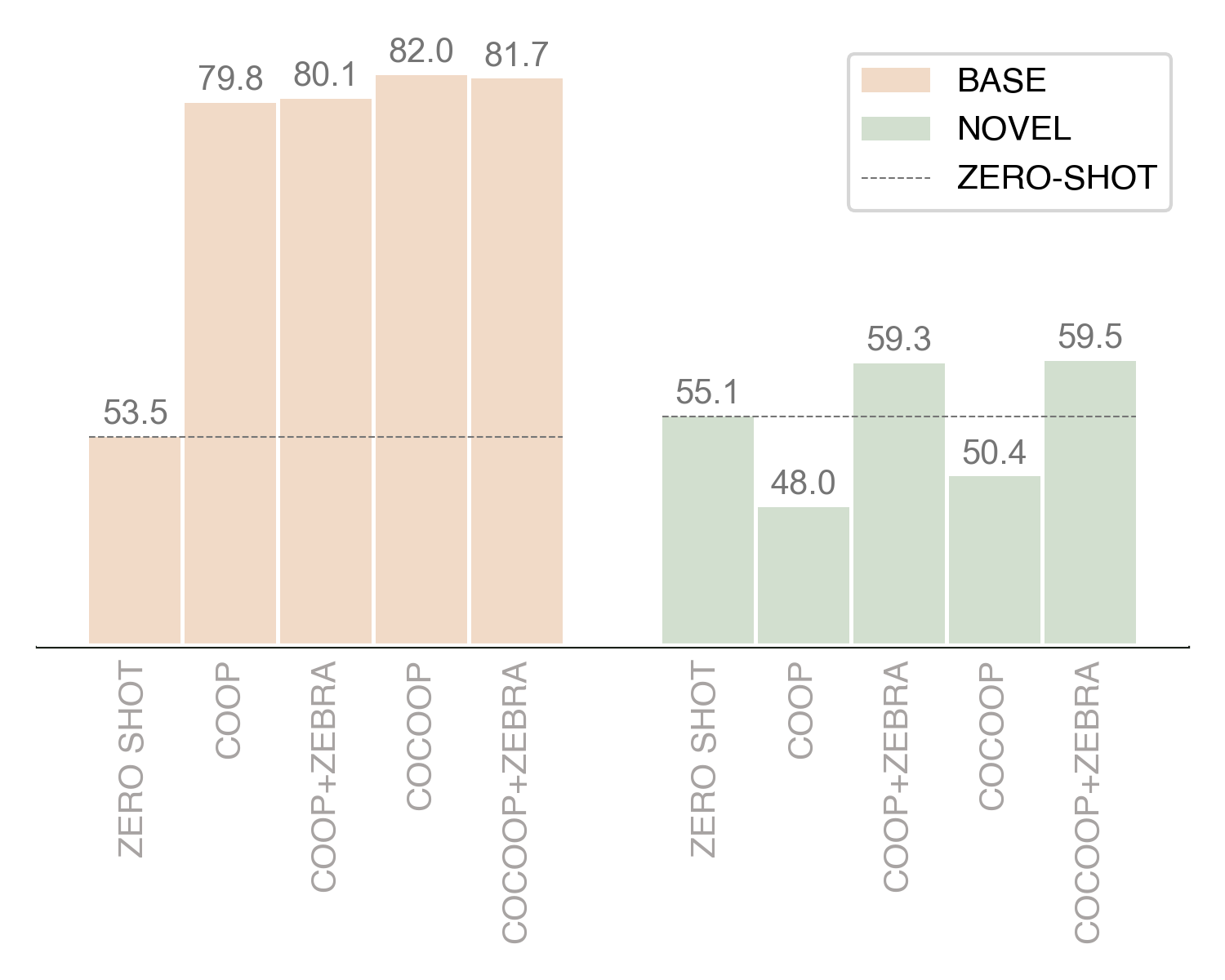}
    \caption{\textbf{Comparison of Base and Novel Performance.} Existing prompt-learning methods improve accuracy on base classes but generalize poorly to novel classes, often performing even worse than zero-shot inference. In contrast, incorporating ZEBRA with these baselines consistently boosts novel-class accuracy while maintaining strong performance on base classes.}
    \label{fig:base_novel_comparison_all}
\end{figure}

\noindent However, in the context of audio-language models, we observe a critical trade-off: while prompt learning consistently improves accuracy on base classes (seen during few-shot training), it often degrades performance on novel classes (unseen during few-shot training). In several cases, prompt learning performs even worse than the original zero-shot model on novel categories (refer to Figure \ref{fig:base_novel_comparison_all} and Table \ref{tab:main_results}). This reveals a fundamental base-to-novel generalization gap under prompt learning. Optimizing prompts solely using base-class supervision encourages overfitting to seen categories, distorting the semantic alignment learned during large-scale pre-training and weakening zero-shot transferability. \\

\begin{figure*}
    \centering
    \includegraphics[width=1.0\linewidth]{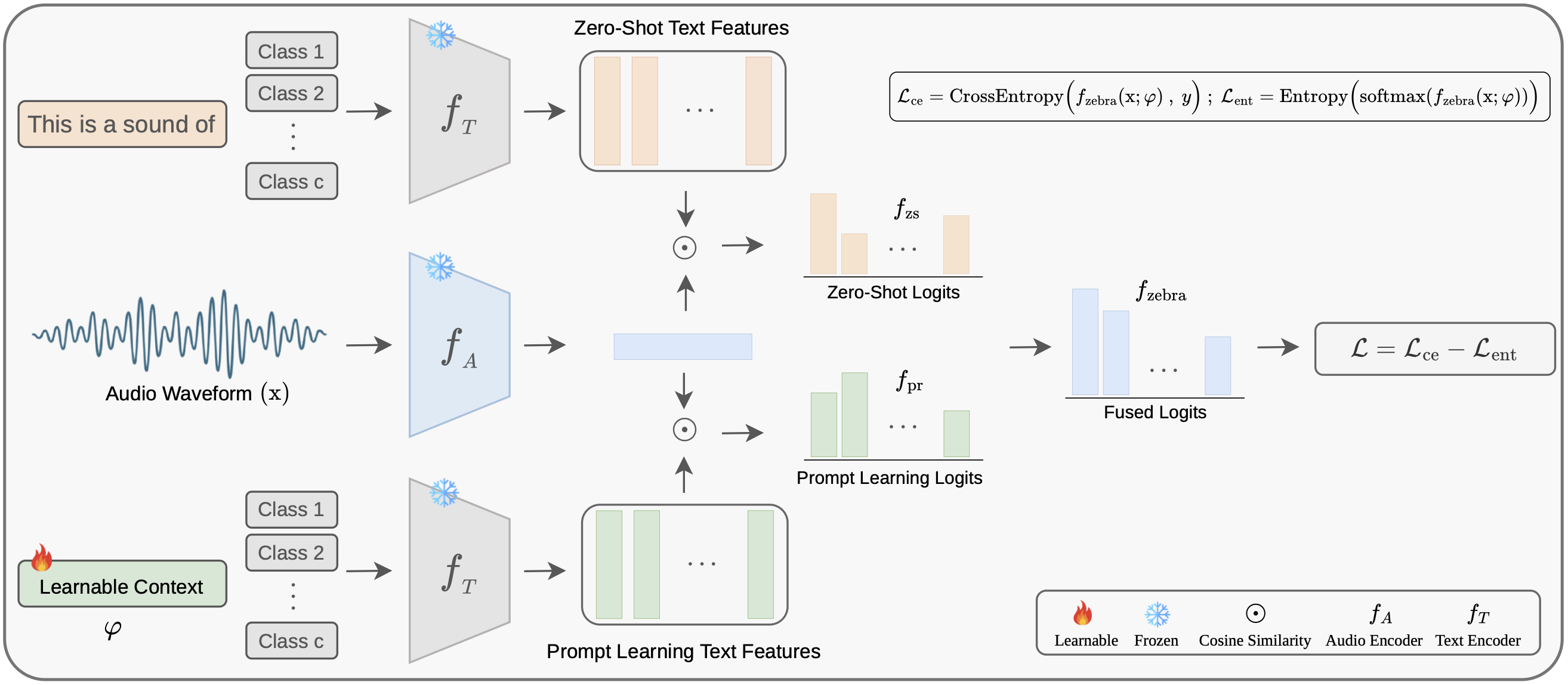}
    \caption{\textsc{ZEBRA} approach operates on top of existing prompt learning methods to bridge the base-to-novel generalization gap, preserving zero-shot transferability while benefiting from supervised adaptation through few-shot prompt learning. ZEBRA introduces no additional learnable parameters to existing prompt learning methods and incurs negligible computational overhead.}
    \label{fig:zebra}
\end{figure*}

\noindent To address this challenge, we propose \textbf{ZEBRA}: \textbf{Z}ero-shot \textbf{E}ntropy-Regularized Prompt Learning for \textbf{B}ase-to-Novel Gene\textbf{RA}lization. ZEBRA is a plug-and-play framework that operates on top of existing prompt learning methods and is designed to preserve zero-shot generalization while still benefiting from supervised adaptation. ZEBRA introduces two complementary mechanisms: First, it fuses zero-shot logits with the prompt-learning logits, effectively anchoring adaptation to the original zero-shot decision space. Second, it introduces a self-entropy regularization term in the loss function to mitigate overconfidence in base classes during few-shot training. By discouraging over-confident predictions, self-entropy prevents excessive specialization to seen categories and promotes smoother decision boundaries, thereby improving generalization to unseen classes. ZEBRA (refer to Figure \ref{fig:zebra} for an overview) is lightweight, as it introduces no additional learnable parameters to existing prompt-learning baselines. The zero-shot logits are computed once and reused during both few-shot training and inference, eliminating extra forward passes through the text encoder and resulting in negligible computational overhead. Our \textit{contributions} are as follows:
\begin{itemize}
    \renewcommand\labelitemi{--}
    \item We identify and analyze a base-to-novel generalization gap in prompt learning for audio-language models, where improvements on base classes often come at the cost of degraded novel-class performance (see Figure \ref{fig:base_novel_comparison_all} or Table ~\ref{tab:main_results}).
    
    \item We propose ZEBRA (Zero-shot Entropy-Regularized Prompt Learning for Base-to-Novel Generalization), a plug-and-play framework that builds upon existing prompt-learning methods without introducing additional learnable parameters and with negligible computational overhead.
    
    \item ZEBRA integrates zero-shot logit fusion to anchor adaptation to the pre-trained decision space and employs self-entropy regularization to reduce overconfidence on base classes, enhancing generalization to unseen categories. 
    
    \item We demonstrate consistent improvements in novel-class generalization across multiple audio classification datasets, significantly reducing the base-to-novel performance gap compared to vanilla baseline prompt learning methods.
\end{itemize}

\section{Related Work}
Inspired by the success of CLIP \cite{radford2021learning} in the image–language domain, many audio–language models have adopted a similar contrastive learning framework. CLAP \cite{elizalde2023clap} and AudioCLIP \cite{guzhov2022audioclip}, for example, extend this paradigm to align audio and textual representations, enabling robust audio classification and cross-modal retrieval. Like CLIP, these models are trained to maximize similarity between matched audio–text pairs while minimizing similarity for mismatched pairs.\\

\noindent Few-shot adaptation has attracted significant attention in vision–language models, particularly for adapting CLIP \cite{gu2023systematic}. Among these, prompt learning has emerged as a parameter-efficient alternative to full fine-tuning, enabling adaptation to downstream tasks by introducing a small set of learnable parameters while keeping the pretrained backbone frozen \cite{zhou2022coop,zhou2022conditional}. This strategy substantially reduces computational and storage costs compared to full model updates. Prompt learning has been successfully applied across natural language processing \cite{liu2023pre}, computer vision \cite{gu2023systematic,hanif2024baple,imam2025noise}, and more recently, the audio-language domain \cite{hanif2024palm,seth2025pat,hanif2025trojanwave}. In audio–language models, approaches such as PALM \cite{hanif2024palm} demonstrate that learning only a small set of prompt vectors can effectively adapt frozen pretrained ALM to achieve strong audio classification performance.

\section{Methodology}
\label{sec:methodology}
\noindent\textbf{Zero-Shot Classification in ALM.} In CLIP-style audio-language models (ALMs)~\cite{radford2021learning}, zero-shot classification is performed by measuring the similarity between the audio representation and a set of class-specific text descriptions. Let $\x$ denote an input audio waveform, and let  \(\txt = \{t_1, t_2, \dots, t_c\}\) represent the set of textual class descriptions for $c$ classes. The prediction logits are computed as:

\begin{equation}
\label{eq:zs_logits}
f_{\mathrm{zs}}(\x) = \bigg\{\operatorname{sim}\big(f_{A}(\x)~,~f_{T}(t_i)\big)\bigg\}_{i=1}^{c},
\end{equation}

\noindent where $f_A$ and $f_T$ denote the audio and text encoders, respectively, and $\operatorname{sim}(\cdot)$ is a cosine similarity function. The predicted label corresponds to the class with the highest similarity score.  \\

\noindent\textbf{Base-to-Novel Generalization.} In the base-to-novel setting, the full label space is defined as $\mathcal{C} = \{1,2,\ldots,c\}$ and is partitioned into base classes $\mathcal{C}_B$ and novel classes $\mathcal{C}_N$, such that $\mathcal{C}_B \cup \mathcal{C}_N = \mathcal{C}$ and $\mathcal{C}_B \cap \mathcal{C}_N = \emptyset$. The model is trained using labeled data from $\mathcal{C}_B$ and is expected to generalize effectively to the unseen (novel) classes in $\mathcal{C}_N$. In the context of prompt learning, the prompts are optimized using base-class samples from the training set, and evaluation is conducted on both base and novel class samples from the test set to assess generalization performance.\\

\noindent\textbf{Prompt Learning in ALM.} 
Textual class descriptions play a central role in zero-shot inference for ALMs; however, manually designed prompts often introduce performance variability and sensitivity to wording. Prompt learning mitigates this issue by introducing auxiliary learnable parameters $\varphi$ into the text encoder $f_T$. These parameters are optimized in a few-shot setting to adapt the frozen pre-trained model to downstream tasks while reducing reliance on manual prompt engineering~\cite{liu2023pre}. Formally, prompt learning learns $\varphi$ by minimizing a task-specific objective over a few-shot training dataset $\mathcal{D}$:
\begin{equation}
\label{eq:pl_logits}
\underset{\varphi}{\operatorname{minimize}} \sum_{(\mathbf{x},y)\in\mathcal{D}}
\mathcal{L}\big(f_{\mathrm{pr}}(\mathbf{x};\varphi), y\big),
\end{equation}
where $(\mathbf{x}, y)$ denotes an audio-label pair from $\mathcal{D}$, $f_{\mathrm{pr}}(\mathbf{x};\varphi)$ represents the model’s prediction conditioned on the learnable prompt parameters $\varphi$, and $\mathcal{L}(\cdot)$ is the task-specific loss function, typically the cross-entropy loss. Two representative instances of  $f_{\mathrm{pr}}(\mathbf{x};\varphi)$ are COOP \cite{zhou2022coop} and COCOOP \cite{zhou2022conditional}.
In COOP, a set of shared learnable context tokens is optimized and prepended to each class name before being passed to text encoder. In COCOOP, the context tokens are dynamically generated conditioned on the audio feature vector, enabling instance-specific adaptation. Although these approaches improve performance on base classes, they optimize prompts solely using base-class supervision and do not explicitly preserve the original zero-shot alignment. As a result, the adapted representation space can become biased toward seen categories, leading to degraded performance on novel classes, as shown in Figure \ref{fig:base_novel_comparison_all}.\\

\noindent\textbf{ZEBRA: Zero-Shot Entropy-Regularized Base-to-Novel Generalization.} 
To mitigate the base-to-novel generalization gap, we propose ZEBRA, a plug-and-play framework that operates on top of existing prompt learning methods. ZEBRA preserves zero-shot transferability while benefiting from supervised adaptation through two complementary mechanisms.\\

\noindent\textit{(i) Zero-shot Logit Fusion:}
Instead of relying solely on prompt-learning logits, we fuse them with the original zero-shot logits during both few-shot training and inference:
\begin{equation}
    \label{eq:zebra_logits}
    f_{\mathrm{zebra}}(\x;\varphi) = \lambda_{\mathrm{zs}}\cdot f_{\mathrm{zs}}(\x) + \lambda_{\mathrm{pr}}\cdot f_{\mathrm{pr}}(\x;\varphi),
\end{equation}
where $f_{\mathrm{zs}} \in \mathbb{R}^{c}$ denotes logits from zero-shot setting, $f_{\mathrm{pr}} \in \mathbb{R}^{c}$ denotes logits from any prompt-learning method (e.g., COOP or COCOOP) and $\lambda_{\mathrm{zs}},\lambda_{\mathrm{pr}}$ control the contribution of the zero-shot and prompt learning components, respectively. This fusion anchors adaptation to the pre-trained decision space.\\

\noindent\textit{(ii) Self-Entropy Regularization:}
To prevent overconfidence on base classes, we introduce a self-entropy regularization term:
\begin{equation}
\mathcal{L}_{\text{ent}} 
=
- \sum_{i} \mathbf{p}^{i}(\mathbf{x}) \log \mathbf{p}^{i}(\mathbf{x}),
\end{equation}
where $\mathbf{p}(\mathbf{x})$ is the softmax probability derived from $f_{\mathrm{zebra}}(\x;\varphi) \in \mathbb{R}^{c}$. 
The final training objective becomes:
\begin{equation}
\nonumber
\mathcal{L}_{\mathrm{zebra}}\big(\x,y;\varphi\big) = \mathcal{L}_{\mathrm{ce}}\big(f_{\mathrm{zebra}},y\big) - \mathcal{L}_{\mathrm{ent}}(\mathrm{softmax}(f_{\mathrm{zebra}})),
\end{equation} 
where $\mathcal{L}_{\mathrm{ce}}(\cdot)$ is the cross-entropy loss. By combining zero-shot logit fusion with entropy regularization, ZEBRA reduces overfitting to base classes while preserving the semantic alignment necessary for robust novel-class generalization. Formally, we optimize following objective during few-shot training on base classes:
\begin{equation}
\label{eq:zebra_objective}
\underset{ \varphi }{\operatorname{minimize}}~~ \sum_{(\x,y)\in\mathcal{D}} \mathcal{L}_{\mathrm{zebra}}\big(\x,y;\varphi\big).
\end{equation} 
During optimization, we minimize the cross-entropy loss while maximizing the self-entropy term, which constructively counteracts overconfident predictions, thereby discouraging overconfidence and reducing overfitting to the base classes. At inference time, predictions are computed from the fused zero-shot and prompt-learning logits, without applying entropy regularization. The predicted label $\hat{y}$ is obtained as follows:
\begin{equation}
\hat{y} = \underset{ i\in \{1,2,\dots,c\} }{\mathrm{argmax}} ~~~ f_{\mathrm{zebra}}^{i}(\x;\varphi),
\end{equation}
where $f_{\mathrm{zebra}}(\x;\varphi) \in \mathbb{R}^{c}$ denotes the fused logit vector over all $c$ classes, and $f_{\mathrm{zebra}}^{i}(\x;\varphi)$ represents the logit corresponding to $i_{\mathrm{th}}$ class. It should be noted that few-shot training is performed using classes in $\mathcal{C}_B$, while evaluation is conducted separately on base classes $\mathcal{C}_B$ and novel classes $\mathcal{C}_N$.\\

\noindent It is worth noting that ZEBRA introduces no additional learnable parameters to existing prompt-learning baselines. The zero-shot logits are computed only once and can be reused during both few-shot training and inference, eliminating the need for additional forward passes through the text encoder. As a result, ZEBRA is a lightweight approach that incurs negligible computational overhead. For an overview of the proposed approach, see Figure~\ref{fig:zebra}.

\section{Experiments and Results}
\noindent \textbf{Models and Datasets.}
For the CLIP-style audio-language backbone, we adopt Pengi \cite{deshmukh2023pengi}, a generative audio-language model consisting of audio and text encoders followed by an LLM decoder. Following the setup of PALM \cite{hanif2024palm}, we discard the decoder and utilize only the pretrained audio and text encoders, effectively employing PENGI in a CLIP-like contrastive framework to leverage its strong pretrained representations and zero-shot generalization capability. We evaluate our method on a diverse set of speech and audio processing tasks, including instrument classification, sound event classification, emotion recognition, vocal sound classification, surveillance sound event classification, acoustic scene classification, and music analysis. Instrument classification is assessed using the Beijing Opera \cite{tian2014study} and NS-Instruments \cite{nsynth2017} datasets. For sound event classification, we use ESC-50 and its ESC50-Actions subset \cite{piczak2015dataset}, as well as UrbanSound8K \cite{salamon2014dataset}. Emotion recognition is evaluated on CREMA-D \cite{cao2014crema} and RAVDESS \cite{livingstone2018ryerson}. Vocal sound classification is conducted using VocalSound \cite{gong_psla}, while surveillance sound event classification relies on SESA \cite{spadini2019sound}. Acoustic scene classification is performed on TUT2017 \cite{heittola2017tut}, and music analysis is evaluated using GT-Music-Genre \cite{sturm2012analysis}. \\

\begin{table*}[!th]
    \centering
    \setlength{\tabcolsep}{3pt}
    \caption{\textbf{Comparison of ZEBRA with Baseline Methods.} 
    While baseline prompt learning methods improve performance on base classes, they often lead to degraded accuracy on novel classes when compared with zero-shot. 
    By incorporating ZEBRA, performance on novel classes is consistently improved while maintaining strong base-class accuracy. 
    Values marked with \textcolor{ForestGreen}{$\blacktriangle$}/\textcolor{BrickRed}{$\blacktriangledown$} denote the increase/decrease in accuracy with respect to the zero-shot performance for the corresponding dataset in each row.
    }
    \scalebox{0.85}{%
    \begin{tabular}{l|cc|cc|cc|cc|cc}
    \toprule
    $\bm{\mathrm{METHODS}}$ $\rightarrow$ & \multicolumn{2}{|c}{$\bm{\mathrm{ZERO~SHOT}}$} & \multicolumn{2}{|c}{$\bm{\mathrm{COOP}}$} & \multicolumn{2}{|c}{$\bm{\mathrm{COCOOP}}$} & \multicolumn{2}{|c}{$\bm{\mathrm{COOP+ZEBRA}}$} & \multicolumn{2}{|c}{$\bm{\mathrm{COCOOP+ZEBRA}}$} \\
    \cmidrule(lr{3pt}){2-3} \cmidrule(lr{3pt}){4-5} \cmidrule(lr{3pt}){6-7} \cmidrule(lr{3pt}){8-9} \cmidrule(lr{3pt}){10-11} 
    $\bm{\mathrm{DATASETS}}$ $\downarrow$ & BASE & NOVEL & BASE & NOVEL & BASE & NOVEL & BASE & NOVEL & BASE & NOVEL  \\
    \midrule \midrule
    Beijing-Opera & 52.00 & 43.48 & 96.10 \uptri{44.1} & 60.87 \uptri{17.3} & 96.00 \uptri{44.0} & 60.88 \uptri{17.3} & 96.03 \uptri{44.0} & 82.61 \uptri{39.1} & 96.20 \uptri{44.2} & 78.26 \uptri{34.7} \\
    CREMA-D & 66.13 & 25.99 & 59.16 \downtri{6.97} & 32.61 \uptri{6.62} & 63.11 \downtri{3.02} & 14.84 \downtri{11.1} & 61.72 \downtri{4.41} & 18.24 \downtri{7.75} & 54.29 \downtri{11.8} & 19.94 \downtri{6.05} \\
    ESC50-Actions & 67.50 & 75.00 & 100.0 \uptri{32.5} & 72.50 \downtri{2.50} & 100.0 \uptri{32.5} & 62.50 \downtri{12.5} & 95.00 \uptri{27.5} & 77.50 \uptri{2.50} & 97.50 \uptri{30.0} & 77.50 \uptri{2.50} \\
    ESC50 & 58.50 & 67.00 & 94.50 \uptri{36.0} & 54.50 \downtri{12.5} & 95.00 \uptri{36.5} & 63.00 \downtri{4.00} & 95.50 \uptri{37.0} & 65.00 \downtri{2.00} & 94.50 \uptri{36.0} & 61.50 \downtri{5.50} \\
    GT-Music-Genre & 56.86 & 36.73 & 76.47 \uptri{19.6} & 53.06 \uptri{16.3} & 83.33 \uptri{26.4} & 45.92 \uptri{9.19} & 74.51 \uptri{17.6} & 52.04 \uptri{15.3} & 83.33 \uptri{26.4} & 37.76 \uptri{1.03} \\
    NS-Instruments & 53.61 & 53.87 & 65.79 \uptri{12.1} & 39.22 \downtri{14.6} & 66.67 \uptri{13.0} & 68.57 \uptri{14.7} & 70.78 \uptri{17.1} & 54.93 \uptri{1.06} & 68.34 \uptri{14.7} & 63.73 \uptri{9.86} \\
    RAVDESS & 23.25 & 38.78 & 59.21 \uptri{35.9} & 32.70 \downtri{6.08} & 60.53 \uptri{37.2} & 40.68 \uptri{1.90} & 59.65 \uptri{36.4} & 42.97 \uptri{4.19} & 60.09 \uptri{36.8} & 43.73 \uptri{4.95} \\
    SESA & 60.00 & 93.33 & 95.56 \uptri{35.5} & 76.67 \downtri{16.6} & 91.11 \uptri{31.1} & 93.33 \uptri{0.00} & 91.11 \uptri{31.1} & 98.33 \uptri{5.00} & 93.33 \uptri{33.3} & 95.00 \uptri{1.67} \\
    TUT2017 & 33.33 & 30.72 & 67.81 \uptri{34.4} & 15.86 \downtri{14.8} & 80.14 \uptri{46.8} & 18.88 \downtri{11.8} & 71.69 \uptri{38.3} & 31.53 \uptri{0.81} & 80.37 \uptri{47.0} & 37.35 \uptri{6.63} \\
    UrbanSound8K & 63.77 & 67.55 & 88.11 \uptri{24.3} & 36.12 \downtri{31.4} & 88.80 \uptri{25.0} & 47.48 \downtri{20.0} & 85.26 \uptri{21.4} & 65.94 \downtri{1.61} & 87.89 \uptri{24.1} & 62.96 \downtri{4.59} \\
    VocalSound & 53.90 & 74.54 & 75.50 \uptri{21.6} & 54.43 \downtri{20.1} & 77.95 \uptri{24.0} & 38.77 \downtri{35.7} & 80.73 \uptri{26.8} & 64.07 \downtri{10.4} & 83.63 \uptri{29.7} & 76.77 \uptri{2.23} \\
    \midrule
    AVERAGE & 53.53 & 55.18 & 79.82 \uptri{26.2} & 48.04 \downtri{7.13} & 82.05 \uptri{28.5} & 50.44 \downtri{4.74} & 80.17 \uptri{26.6} & 59.37 \uptri{4.19} & 81.75 \uptri{28.2} & 59.50 \uptri{4.31} \\
    \bottomrule
    \end{tabular}
    }
    \label{tab:main_results}
\end{table*}
\begin{table}[!h]
    \centering
    \setlength{\tabcolsep}{6pt}
    \caption{Ablation on zero-shot logits and entropy loss.
    }
    \scalebox{0.85}{%
    \begin{tabular}{cc|cc}
    \toprule
    $\bm{\mathrm{ZERO{-}SHOT}}$ &  $\bm{\mathrm{ENTROPY}}$ & $\bm{\mathrm{BASE}}$ & $\bm{\mathrm{NOVEL}}$\\
    \midrule
    \xmark & \xmark & 79.82 & 48.04 \\
    \cmark & \xmark & 80.31 & 58.83 \\
    \xmark & \cmark & 81.21 & 48.10 \\
    \cmark & \cmark & 80.17 & 59.37\\
    \bottomrule
    \end{tabular}
    }
    \label{tab:ablation_zs_ent}
\end{table}
\begin{table}[!h]
    \centering
    \setlength{\tabcolsep}{4pt}
    \caption{Runtime and ECE of baselines with and without \textsc{ZEBRA}. $(\dagger)$ indicates results obtained using \textsc{ZEBRA}.}
    \scalebox{0.82}{%
    \begin{tabular}{l|cccccc}
    \toprule
    $\bm{\mathrm{METHODS}}$ & $\bm{\mathrm{COOP}}$ & $\bm{\mathrm{COOP}^{\dagger}}$ & $\bm{\mathrm{COCOOP}}$ & $\bm{\mathrm{COCOOP}^{\dagger}}$ \\
    \midrule
    ${\mathrm{TIME~({Train})}}$ & $\mathrm{19m~36s}$ & $\mathrm{19m~45s}$ & $\mathrm{45m~37s}$ & $\mathrm{45m~41s}$  \\
    ${\mathrm{TIME~({Test})}}$ & $\phantom{0}\mathrm{2m~40s}$ & $\phantom{0}\mathrm{2m~42s}$ & $\phantom{0}\mathrm{3m~37s}$ & $\phantom{0}\mathrm{3m~38s}$\\
    \midrule 
    $\mathrm{ECE~(BASE)}$  & $0.0677$ & $0.0755$ & $0.0821$ & $0.0777$  \\
    $\mathrm{ECE~(NOVEL)}$ & $0.2033$ & $0.1997$ & $0.2738$ & $0.2253$ \\
    \bottomrule
    \end{tabular}
    }
    \label{tab:computational_cost}
\end{table}

\noindent \textbf{Baseline Methods.}
We consider ZERO-SHOT, COOP \cite{zhou2022coop}, and COCOOP \cite{zhou2022conditional} as baselines. COOP and COCOOP are prompt learning methods originally proposed for vision-language models that replace handcrafted prompts with learnable context tokens optimized in the text encoder’s input embedding space; COCOOP further introduces instance-conditioned prompts via feedback from the audio encoder, whereas COOP learns context tokens shared across instances. We adopt their audio-language adaptations as implemented in PALM \cite{hanif2024palm}, where the vision encoder is replaced with an audio encoder. We exclude PALM itself as a baseline because its class-specific learnable vectors limit base-to-novel generalization, while COOP and COCOOP are class-agnostic and can generalize to unseen classes. Our method builds upon COOP and COCOOP in a plug-and-play manner, seamlessly integrating with their learned prompt representations without introducing additional trainable parameters or requiring further hyperparameter tuning. It operates on top of the existing framework, preserving their training protocols and generalization capabilities, while providing consistent improvements with minimal computational overhead and no modifications to the underlying model architecture.\\

\noindent \textbf{Implementation Details.} Except for ZERO-SHOT, all methods are trained for $50$ epochs under the standard few-shot setting, using $16$ randomly sampled training examples per \textit{base} class. Inference is conducted on the full test set using the predefined \textit{base–novel} class split. Few-shot training is done with stochastic gradient descent (SGD) at a learning rate of $0.05$, and performance is measured in terms of accuracy. Each method is evaluated with three different random seeds, and we report the average results. For ZERO-SHOT, we adopt the default prompt template: \texttt{This is a recording of \{CLASS NAME\}}, for a fair comparison. We run all experiments using \texttt{NVIDIA RTX A6000} GPU. For the contribution weights in Equation~\ref{eq:zebra_logits}, we empirically set $\lambda_{\mathrm{zs}} = 0.5$ and $\lambda_{\mathrm{pr}} = 0.5$. We also scale the $\mathcal{L}_{\mathrm{ent}}$ by a factor of $0.05$ during few-shot training. \\

\noindent \textbf{Results and  Discussion.} We report a comparison of ZERO-SHOT, prompt-learning methods, and their ZEBRA-enhanced variants in Table~\ref{tab:main_results}. As discussed earlier, baseline methods (COOP and COCOOP) substantially improve performance on base classes compared to ZERO-SHOT (e.g., $+26.2\%$ and $+28.5\%$, respectively). However, they generalize poorly to novel classes, exhibiting notable drops in accuracy (e.g., $-7.13\%$ and $-4.74\%$, respectively) and, in both cases, performing below the ZERO-SHOT novel-class average of $55.18\%$. On the other hand, ZEBRA enhances their novel-class performance by $+4.19\%$ and $+4.31\%$, respectively, bringing the accuracy above the ZERO-SHOT novel-class average of $55.18\%$ while maintaining competitive performance on base classes. Results demonstrate that, on average, ZEBRA improves novel-class generalization while maintaining base performance, achieving these gains with no additional parameters and negligible computational overhead (see Table \ref{tab:computational_cost}).\\

\noindent \textbf{Ablative Analysis.} We analyze the impact of the fusion of zero-shot logits and self-entropy loss term on ZEBRA’s performance in Table~\ref{tab:ablation_zs_ent} (using the COOP baseline). The results indicate that most of the improvement comes from incorporating zero-shot logits, while the self-entropy term provides a further, albeit marginal, gain. We also report runtime (training and testing, averaged across all datasets) and Expected Calibration Error (ECE) in Table~\ref{tab:computational_cost}. ZEBRA introduces negligible computational overhead while consistently reducing ECE on average across both base and novel classes, demonstrating improved calibration without sacrificing efficiency.

\section{Conclusion}
We introduced ZEBRA, a lightweight, plug-and-play framework for improving base-to-novel generalization in audio–language models. While existing prompt-learning methods substantially boost base-class performance, they often suffer from degraded generalization to novel classes, frequently underperforming the zero-shot baseline. ZEBRA effectively mitigates this trade-off by leveraging zero-shot knowledge and entropy-regularized logits, consistently improving novel-class accuracy while preserving strong base-class performance. ZEBRA adds no learnable parameters and incurs negligible computational overhead, making it a simple yet effective enhancement to existing prompt-learning methods.

\section{Generative AI Use Disclosure}
We confirm that an LLM was used solely for writing refinement (grammar, wording, and clarity). All ideas, analyses, and conclusions are the authors’ own.

\bibliographystyle{IEEEtran}
\bibliography{main}

\end{document}